\documentclass[%
reprint,
superscriptaddress,
showpacs,preprintnumbers,
 amsmath,amssymb,
aps,
pre,
floatfix%,
]{revtex4-1}

\usepackage{float}
\usepackage{graphicx}% Include figure files
\usepackage{dcolumn}% Align table columns on decimal point
\usepackage{bm}% bold math
\usepackage{bbold}
\usepackage[multidot]{grffile}
\usepackage[colorlinks=true,allcolors=blue, breaklinks=true]{hyperref}% add hypertext capabilities
\newcommand{\vect}[1]{\bm{\mathrm{#1}}}

\DeclareMathOperator{\dd}{d\!} % pour avoir un d de dérivée droit

\usepackage{xcolor}

\usepackage{xspace}

\usepackage[utf8]{inputenc}

\begin{document}

%\preprint{APS/123-QED}

\title{Nonautonomous driving induces stability in network of identical oscillators}

\author{Maxime Lucas}
\email{m.lucas@lancaster.ac.uk}
\affiliation{Department of Physics, Lancaster University, Lancaster LA1 4YB, United Kingdom}
\affiliation{Dipartimento di Fisica e Astronomia, Universit{\`a} di Firenze, INFN and CSDC, Via Sansone 1, 50019 Sesto Fiorentino, Firenze, Italy}
\author{Duccio Fanelli}
\email{duccio.fanelli@gmail.com}
\affiliation{Dipartimento di Fisica e Astronomia, Universit{\`a} di Firenze, INFN and CSDC, Via Sansone 1, 50019 Sesto Fiorentino, Firenze, Italy}
%
%\author{Julian Newman}
%\email{j.newman1@lancaster.ac.uk}
%\affiliation{Department of Physics, Lancaster University, Lancaster LA1 4YB, United Kingdom}
%
\author{Aneta Stefanovska}
\email{aneta@lancaster.ac.uk}
\affiliation{Department of Physics, Lancaster University, Lancaster LA1 4YB, United Kingdom}

\date{\today}% It is always \today, today,
             %  but any date may be explicitly specified

\begin{abstract}
Nonautonomous driving of an oscillator has been shown to enlarge the Arnold tongue in parameter space, but little is known about the analogous effect for a network of oscillators. To test the hypothesis that deterministic nonautonomous perturbation is a good candidate for stabilising complex dynamics, we consider a network of identical phase oscillators driven by an oscillator with a slowly time-varying frequency. We investigate both the short- and long-term stability of the synchronous solutions of this nonautonomous system. For attractive couplings we show that the region of stability grows as the amplitude of the frequency modulation is increased, through the birth of an intermittent synchronisation regime. For repulsive couplings, we propose a control strategy to stabilise the dynamics by altering very slightly the network topology. We also show how, without changing the topology, time-variability in the driving frequency can itself stabilise the dynamics. As a by-product of the analysis, we observe chimera-like states. We conclude that  time-variability-induced stability phenomena are also present in networks, reinforcing the idea that this is quite realistic scenario for living systems to use in maintaining their functioning in the face of ongoing perturbations.
\end{abstract}

%\pacs{05.45.Xt, 05.65.+b, 89.75.Fb}% PACS, the Physics and Astronomy
                             % Classification Scheme.
%\keywords{Suggested keywords}%Use showkeys class option if keyword
                              %display desired
\maketitle

\section{Introduction} \label{sec:introduction}

Many natural and man-made systems consist of interacting oscillatory units~\cite{Haken:75,nicolis1977,goldbeter1997}. Such coupled oscillators can yield a plethora of complex dynamical behaviours.  Of utmost importance is the case  of synchronous dynamics, when all oscillators self-organise and oscillate identically~\cite{pikovsky2003,strogatz2004}. Synchronous dynamics can be crucially needed, such as for the beating of the heart, and can also be severely detrimental, such as in epileptic seizures~\cite{mormann2000,andrzejak2016}. Phenomena such as ageing~\cite{Iatsenko:13a} and an{\ae}stesia~\cite{Stankovski:16} are known to alter synchrony of dynamics. These systems can be divided into two classes, based on the nature of their interactions being either {\em attractive}, where all oscillators tend to be synchronous, i.e.\ have the same phase, or {\em repulsive}, where the oscillators tend to be not synchronous \cite{Stankovski2017}.

The Kuramoto model~\cite{kuramoto1984} has served as a paradigmatic model for such systems, and many later works have extended the theory of synchronisation to networks with complex topologies~\cite{arenas2008,dorfler2014,rodrigues2016}. Coupled oscillator models have been applied successfully in diverse areas of the sciences~\cite{goldbeter1997,Glass2001}, including circadian rhythms~\cite{leloup1999}, cardio-respiratory dynamics~\cite{Stefanovska2000,Stefanovska:01a,suprunenko2013}, and metabolic oscillations~\cite{lancaster2016}.

Beyond the internal dynamics, many systems are subject to ongoing external influence from their environment. That external influence drives the system, and its action can, in general, change with time. The day-light cycles drive the suprachiasmatic nucleus  neurons in our brain, which in turn control all circadian rhythms in our bodies, making sure the body's different clocks tick together~\cite{flourakis2015}. Day-light cycles are subject to seasonality. Moreover, the neurons's firing is influenced by intra and extracellular ionic concentrations, which in turn affects synchronisation~\cite{pfeuty2003}. In some cases, like for fish communities in Japan~\cite{ushio2018}, the time-variability of the system yields time-varying properties and stability~\cite{bick2011,atsumi2012}, which was found crucial to the functioning of the community. Moreover, in real-world systems, short-term properties are sometimes more physically relevant than the asymptotic or long-term ones~\cite{newman2018}. Non-driven networks with time-varying parameters have started to attract attention recently~\cite{petkoski2012,pietras2016,petit2017,lu2018,lucas2018b}. Driven networks of identical oscillators have been studied theoretically in specific cases, such as regular topologies~\cite{radicchi2006}, or hierarchical networks with only a few oscillators being driven~\cite{kori2004, kori2006}.

However, to date, those few studies either considered fixed-frequency driving, or modelled the time-variability statistically as just noise. Here, we consider a deterministically but aperiodically time-varying frequency of the external driving, playing the role of the ever changing environment. We bring insight about the effect of the time-variability on the stability -- both short- and long-term -- of synchronous solutions, analytically and numerically. Surprisingly, we show that adding time-variability in the driving can increase the stability of the driven network. In the attractive coupling case, we generalise results from \cite{lucas2018, newman2018} to networks. Weakly coupled non-linear oscillators with slowly varying parameters can be reduced to a phase model \cite{kurebayashi2013,kurebayashi2015,park2016}, which suggests that our results apply to many other types of oscillators. Finally, our results are valid for networks of any size and any topology.

The paper is organized as follows. In Sec.~\ref{sec:model}, we introduce our driven network of oscillators model. In Sec.~\ref{sec:analysis}, we provide a general linear stability analysis of the synchronous solutions. First, in Sec.~\ref{sec:a}, the analysis is performed for the fixed-frequency driving case. Then, in Sec.~\ref{sec:na}, the analysis is extended to the general nonautonomous case, where the driving frequency is time-dependent. These results are then discussed and confirmed numerically in Sec.~\ref{sec:stability} for the two possible cases. For the attractive and repulsive cases, in Sec.~\ref{sec:attractive} and Sec.~\ref{sec:repulsive} respectively, we show how the stability region is enlarged and describe short-time properties. Furthermore, for the repulsive case, in Sec.~\ref{sec:repulsive}, we evidence a link between topology and dynamics, and suggest a control strategy based on it. Moreover, we show how time-variability can counter-balance the desynchronising effect of topology and enhance synchrony. As a by-product, we observe chimera-like state. Finally, in Sec.~\ref{sec:conclusion} we conclude with a brief summary of the results.

\section{Model}
\label{sec:model}

We consider a driven network of $N$ identical oscillators defined by phases $\theta_i$ and frequency $\omega$,
\begin{equation}
\dot \theta_i = \omega + D \sum_{j=1}^N A_{ij} \sin (\theta_i - \theta_j) + \gamma \sin [ \theta_i - \theta_0(t)] ,
\label{eq:system}
\end{equation}
for $i=1,\dots,N$, with coupling constant $D$, and where $\vect{A}$ stands for any (undirected) adjacency matrix with elements $A_{ij} \in \{0,1\}$. Each oscillator is driven with strength $\gamma \ge 0$ by the same external oscillator with phase $\theta_0(t)$ and time-varying frequency
\begin{equation}
\dot \theta_0 = \omega_0 [1 + k f(\omega_m t)],
\label{eq:driving}
\end{equation}
where $\omega_0$ is the non-modulated frequency, $f$ is a bounded function, and $k$ and $\omega_m$ are the amplitude and frequency of the imposed modulation, respectively. Note that $f(\omega_m t)$ is a generic function, and need not be periodic; without loss of generality, we bound its image in $[-1, 1]$. System \eqref{eq:system}-\eqref{eq:driving} is a direct generalisation to networks of the single driven oscillator system presented in~\cite{lucas2018}.

The non-driven network, $\gamma=0$, is an autonomous system. In this case, a (fully) \textit{synchronous} solution always exists, i.e. a solution where all oscillators are in the same state at all times, $\theta_i = \theta_j$ for all $i \neq j$. This is as long as the network is connected, meaning that there is a path between any two nodes (oscillators). The synchronous solution is stable (unstable) if the coupling between oscillators is \textit{attractive} (\textit{repulsive}), i.e., $D<0$ ($>0$)~\cite{toenjes2009}. In the repulsive case, the system is attracted to an incoherent state.

\section{Theoretical analysis} \label{sec:analysis}

In the driven system, $\gamma \neq 0$, synchronous solutions also exist. It is convenient to go to the rotating frame of the driving, $\psi_i = \theta_i - \theta_0(t)$, where system~\eqref{eq:system} is rewritten
\begin{equation}
\dot \psi_i = \Delta \omega(t) + D \sum_{j=1}^N A_{ij} \sin(\psi_i - \psi_j) + \gamma \sin \psi_i ,
\label{eq:rot_frame}
\end{equation}
for $i=1,\dots,N$, where $\Delta \omega(t) = \omega - \omega_0 [1 + k f(\omega_m t)]$ is the time-dependent frequency mismatch. We refer to the $\theta_i$ as the \textit{phases}, and to the $\psi_i$ as the \textit{phase differences} between the phase of the driving and that of the $i$-th oscillator. System~\eqref{eq:rot_frame} is nonautonomous due to the time-dependent frequency mismatch, and is hence hard to treat in general. In the rest of the study, we assume $\omega_0$ is modulated slowly, i.e., $\omega_m \ll \omega_0$, and use an adiabatic approach to study the existence and stability of synchronous solutions.

\subsection{Autonomous case} \label{sec:a}

To that end, we first consider the simpler case of a constant driving frequency, recovered for $k=0$, for which $\Delta \omega(t) = \Delta \omega = \omega - \omega_0$ is constant and system~\eqref{eq:rot_frame} is autonomous. A \textit{synchronous} solution, $\psi_i = \tilde \psi$ for all $i=1,\ldots,N$, always exists, and obeys
\begin{equation}
\dot{\tilde{\psi}} = \Delta \omega + \gamma \sin \tilde \psi,
\label{eq:adler}
\end{equation}
which is the so-called Adler equation describing a single driven oscillator~\cite{pikovsky2003}. From Eq.~\eqref{eq:adler}, two types of synchronous solutions can be identified: \textit{synchronised} (to the driving), or not . First, if $\gamma \ge | \Delta \omega |$ is satisfied, there exists a stable fixed point. This fixed point corresponds to the synchronous synchronised (SS) solution, $\psi_{\textsc{ss}} = \pi - \arcsin(-\Delta \omega / \gamma)$, characterised by a constant phase difference, $\dot \psi_{\textsc{ss}} = 0$. Second, if the synchronisation condition is not met, $\gamma < | \Delta \omega |$, there exists a synchronous not synchronised (SNS) type of solutions, denoted $\psi_\textsc{sns}$, which grows (or decays) monotonically, $\dot \psi_{\textsc{sns}} > 0$ (or $<0$).

%The condition for synchronisation is that Eq.~\eqref{eq:adler} has a stable fixed point. Hence, two types of synchronous solutions exist: \textit{synchronised} to the driving, if $\gamma > | \Delta \omega |$, and not synchronised to the driving otherwise. The synchronous synchronised (SS) solution has constant phase difference, $\dot \psi_{\textsc{ss}} = 0$, whereas the phase difference in the synchronous not synchronised (SNS) solution grows (or decays) monotonically, $\dot \psi_{\textsc{sns}} > 0$ (or $<0$).

We now investigate the linear stability against a small heterogeneous perturbation $\delta \psi_i$ around those solutions. This is determined by linearising Eq.~\eqref{eq:rot_frame} for each node around a solution $\tilde \psi(t)$, which stands for either the SS or SNS,
\begin{equation}
\delta \dot \psi_i = - D  \sum_{j=1}^N L_{ij} \delta \psi_j + \gamma \delta \psi_i \cos \tilde \psi(t) .
\label{eq:linearised}
\end{equation}
Here, $L_{ij} = A_{ij} - K_i \delta_{ij}$ denotes the Laplacian matrix of the network, defined in terms of the connectivity of each node $K_i = \sum_{j=1}^N A_{ij}$, and the Kronecker delta $\delta_{ij}$. One can now decouple this $N$-dimensional problem by projecting it onto the eigenbasis of the Laplacian, defined as
%
%\begin{equation}
$\sum_{j=1}^N L_{ij} \phi_i^{\alpha} = \Lambda^{\alpha} \phi_i^{\alpha}$,
%\end{equation}
%
where the $\phi^{\alpha}$ are the eigenvectors associated to the eigenvalues $\Lambda^{\alpha}$, for $\alpha=1, \ldots, N$. The latter are non-positive and real, since the network is assumed to be undirected, and ordered $\Lambda^1 = 0 > \Lambda^2 \ge \ldots \ge \Lambda^N$. The perturbation can be decomposed in that basis, and we look for solutions of the form $\delta \psi_i(t) = \sum_{\alpha=1}^{N} c_{\alpha} e^{\int_0^{t} \lambda^{\alpha}(t') \dd t'} \phi_i^{\alpha}$. Plugging into Eq.~\eqref{eq:linearised} and solving for each $\alpha$ yields the instantaneous Lyapunov exponent (LE) spectrum
\begin{equation}
\lambda^{\alpha}(t) = -D \Lambda^{\alpha} + \gamma \cos \tilde \psi(t) ,
\label{eq:insta_lambda}
\end{equation}
which is completely general and also valid in the nonautonomous case, as will be seen later. Now in the autonomous case, $\tilde \psi(t)$ is periodic modulo $2\pi$ \cite{pikovsky2003} and hence (long-term) Lyapunov exponent spectrum is well defined as the time-average of the instantaneous values
\begin{equation}
\lambda^{\alpha} = -D \Lambda^{\alpha} + \gamma \langle \cos \tilde \psi(t) \rangle .
\label{eq:lambda}
\end{equation}
An explicit form of formula~\eqref{eq:lambda} is obtained by splitting it into two cases
\begin{equation}
\lambda^{\alpha} =
\left\{
\begin{array}{ll}
- D \Lambda^{\alpha} - \sqrt{\gamma^2 - \Delta \omega^2} & \text{if } \gamma > | \Delta \omega | , \\
- D \Lambda^{\alpha} & \text{else} . \\
\end{array}
\right.
\label{eq:LES_cases_a}
\end{equation}
For the synchronised case, $\gamma > | \Delta \omega |$, the explicit solution $\psi_{\textsc{ss}} = \pi - \arcsin(-\Delta \omega / \gamma)$, which is constant,  was plugged into formula~\eqref{eq:lambda}. For the  not synchronised case, the solution is $\psi_{\textsc{sns}}$, for which the averaging term in formula~\eqref{eq:lambda} vanishes.

\subsection{Nonautonomous case} \label{sec:na}

In general, the driving frequency is time-dependent, $k\neq0$, and synchronous solutions obey a nonautonomous version of Eq.~\eqref{eq:rot_frame}
\begin{equation}
\dot{\tilde{\psi}} = \Delta \omega(t) + \gamma \sin \tilde \psi,
\label{eq:adler_na}
\end{equation}
which was studied in~\cite{jensen2002,lucas2018}. As previously mentioned, here and throughout the paper, we assume slow modulation of the frequency, i.e. $\omega_m$ small. Equivalently, $\Delta \omega(t)$ varies much more slowly than the dynamics of the system. Hence, there is a separation of timescales: $\Delta \omega(t)$ is the slow variable, and $\tilde \psi(t)$ the fast one. Thus, over the fast timescale, the frequency mismatch is quasi-static, and the slowly moving point attractor $\psi_{\textsc{ss}}(t) = \pi - \arcsin(-\Delta \omega(t) / \gamma)$ is followed adiabatically, when it exists~\cite{jensen2002}. This state corresponds to the SS solution of the autonomous case, and differs with it in being only \textit{quasi}-constant, i.e., $\dot{\tilde{\psi}}_{\textsc{ss}} = 0$ over the fast timescale, and existing only at times such that $\gamma > | \Delta \omega (t) |$.

\begin{figure}[t]
	\centering
	\includegraphics[width=\linewidth]{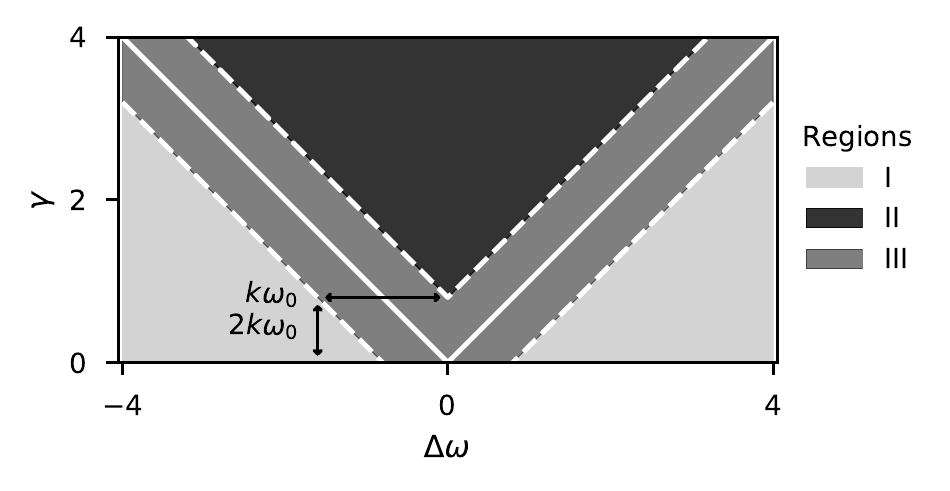}
	%{fig-tongue-theo_D_-0.5_w0_2.00.pdf}
	\caption{Regions of existence of synchronous solutions. Region I: no synchronisation (SNS). Region II: synchronisation (SS). Region III: intermittent synchronisation (SIS).}
	\label{fig:regions}
\end{figure}

Consequently, in contrast to the autonomous case, Eq.~\eqref{eq:adler_na} shows three types of synchronous solutions: two of them correspond to the SS and SNS solutions as discussed in the autonomous case, whereas a third type exhibits \textit{intermittent} synchronisation. Three regions in parameter space can thus be defined, each corresponding to the existence of one of those types solutions~\cite{lucas2018}, as illustrated in Fig.~\ref{fig:regions}. In region I, the condition $\gamma > | \Delta \omega (t) |$ is not met at any time. The solution grows (or decays) monotonically and we denote it $\psi_{\textsc{sns}}(t)$. In region II, the condition $\gamma > | \Delta \omega (t) |$ is met at all times. The solution is denoted by $\psi_{\textsc{ss}}(t)$ and has an approximately null time derivative, as described above. In region III,  $\gamma > | \Delta \omega (t) |$ is met only at certain times. The phase difference alternates between times of growth (or decay), and quasi-constant (bounded) epochs. We call it synchronous intermittently synchronised (SIS), and denote it $\psi_{\textsc{sis}}(t)$. The three types of solutions are illustrated in Fig.~\ref{fig:attr-intermittent}(a) and will be discussed in the next section.

The three regions, depicted in Fig.~\ref{fig:regions}, are equivalently defined by the following time-independent conditions: (I) $\gamma \ge |\Delta \omega| + \omega_0 k$, (II) $\gamma \le |\Delta \omega| - \omega_0 k$, and (III) $|\Delta \omega| - \omega_0 k \le \gamma \le |\Delta \omega| + \omega_0 k$, where we denote $\Delta \omega \equiv \omega~-~\omega_0$ the central frequency mismatch, or alternatively, the frequency mismatch of the autonomous case~\cite{lucas2018}. Note that these regions are defined based on the condition of existence of the aforementioned types of solutions, and not based on their stability.

The adiabatic assumption allows us to obtain the following formula for the adiabatic LE spectrum, as a nonautonomous version of formula~\eqref{eq:LES_cases_a}, by retracing the same reasoning
\begin{equation}
\lambda^{\alpha}(t) =
\left\{
\begin{array}{ll}
- D \Lambda^{\alpha} - \sqrt{\gamma^2 - \Delta \omega(t)^2} & \text{for } t : \gamma > | \Delta \omega (t) | , \\
- D \Lambda^{\alpha} & \text{else} . \\
\end{array}
\right.
\label{eq:instant-lyap}
\end{equation}
Equation~\eqref{eq:instant-lyap} enables us to draw conclusions about the stability of the aforementioned states. The stability of the SS solution can be assessed by the first condition in Eq.~\eqref{eq:instant-lyap}. The stability of the SNS is determined by the second condition in Eq.~\eqref{eq:instant-lyap}. Finally, the stability of the SIS is determined by the first condition of Eq.~\eqref{eq:instant-lyap} at times such that $\gamma > | \Delta \omega (t) |$, and the second condition the rest of the time.

Note that this result holds true regardless of the network size, topology, or the shape of the frequency modulation function $f(\omega_m t)$. 

\section{On the stability of the synchronous solution} \label{sec:stability}

We now examine the stability of the synchronous solutions further via formula~\eqref{eq:instant-lyap} in the attractive and repulsive cases.

\subsection{Attractive case} \label{sec:attractive}

In this case, $D<0$, all oscillators tend to display the same phase, when not driven. For a given network, only the largest LE $\lambda_{max}(t)$ determines the stability. In this case, it is $\lambda_{max}(t) = \lambda^1(t)$ which corresponds to $\Lambda^1=0$ in formula~\eqref{eq:instant-lyap}, and reads
\begin{equation}
\lambda_{max}(t) =
\left\{
\begin{array}{ll}
 - \sqrt{\gamma^2 - \Delta \omega(t)^2} & \text{for } t : \gamma > | \Delta \omega (t) | , \\
0 & \text{else}, \\
\end{array}
\right.
\label{eq:attr-instant-lyap}
\end{equation}
a condition which is identical to that obtained for the single oscillator case considered in~\cite{lucas2018}, even though we have now an arbitrary network of oscillators.

For the sake of clarity, we set $f(\omega_m t) = \sin (\omega_m t)$ in numerical examples. Note, however, that the analysis is independent of the explicit form of $f(\omega_m t)$, which can in general be aperiodic, or even defined only over a finite timespan, as discussed in~\cite{newman2018}.

\begin{figure}[b!]
	\centering
	\includegraphics[width=\linewidth]{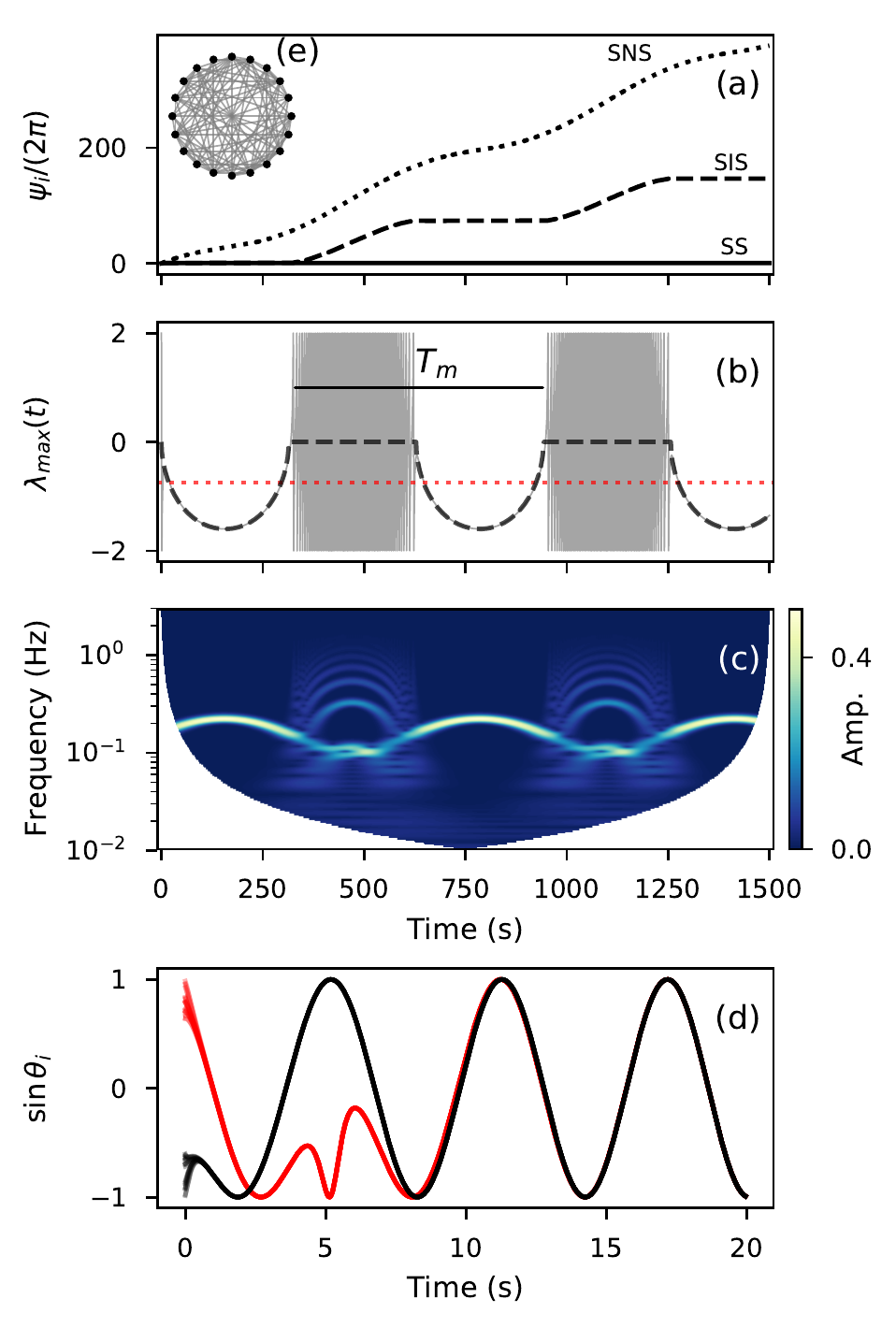}
	%{net_psi_random_N_20_D_-0.5_gammas_[1. 2. 3.]_dw_2.00_w0_1.00_k_0.4_ic_randsync-lys-combined4}
	\caption{Finite-time dynamics and stability, in the attractive case $D<0$. (a) Typical dynamics of the phase difference $\psi_i$ in the three regimes: synchronised (solid, $\gamma=3$), intermittently synchronised (dashed, $\gamma=2$), and not synchronised (dotted, $\gamma=1$). Panels (b)-(d) only consider the intermittently synchronised trajectory. (b) Instantaneous stability, measured by the instantaneous LE (grey) given by Eq.~\eqref{eq:insta_lambda} and the adiabatic approach (dashed black) of Eq.~\eqref{eq:attr-instant-lyap}. The average LE, as defined in Eq.~\eqref{eq:attr-avg-lyap}, is negative (dotted red). (c) Time-frequency representation computed applying continuous Morlet wavelet transform with central frequency 3 to the signal $y(t)=\sin \theta_1(t)$. Epochs of boundedness (unbounded drift) $\psi_i$ correspond to negative (zero) $\lambda(t)$, and (no) frequency entrainment in (c) (respectively). (d) Convergence of two different initial conditions, in red and black, respectively, in the intermittent synchronisation regime. Each initial condition is a quasi-synchronous $N$-dimensional state. Each state first quickly becomes synchronous, and then converges to the synchronised solution. Parameters are $N=20$, $D=-0.5$,  $k=0.4$, $\Delta \omega=2$, $\omega_0=1$. (e) Network used for the numerics: random with connection probability $p=0.5$ and $\Lambda^N=-15.9$. Note that the results do not depend on the network considered.}
	\label{fig:attr-intermittent}
\end{figure}

\begin{figure*}[thb!]
	\centering
	\includegraphics[width=\linewidth]{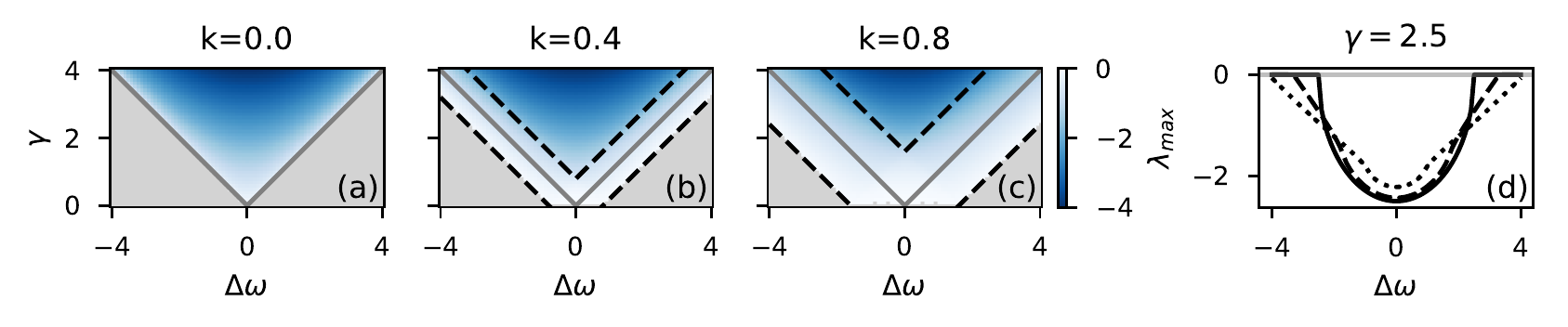}
	%{fig-tongues-attr_D_-0.5_w0_2.00_ks_[0.0, 0.4, 0.8]-horizontal4slice}
	%{../fig-tongues-attr_D_-0.5_w0_2.00_ks_[0.0, 0.4, 0.8]-horizontal.pdf}
	%{Fig1.pdf}
	%{fig-tongues-attr_D_-0.5_w0_2.00_ks_[0.0, 0.4, 0.8]-horizontal4}
	\caption{Stability region increases with amplitude of non-autonomicity $k$, in the attractive case $D<0$. (a)-(c) Quantitative stability measured by the LE of formula~\eqref{eq:attr-avg-lyap}: the region of stability (shades of blue) increases with $k$. Gray values represent zero values of the LE. (d) Same LE values as in (a)-(c), shown only for $\gamma=2.5$, for the different $k$ values: $0.0$ (solid), $0.4$ (dashed), and $0.8$ (dotted). The range of $\Delta \omega$ for which $\lambda_{max}<0$ increases with $k$. Parameters are $D=-0.5$, $w_0=2$. This picture holds true for any network topology.}
	\label{fig:fig-tongues-attr}
\end{figure*}

Typical dynamics of the $\psi_i$ is shown for the three regions in Fig.~\ref{fig:attr-intermittent}(a). The SS (solid) and SNS (dotted) are bounded and unbounded, respectively, as mentioned in the previous section. For the SS (SNS), $\gamma > | \Delta \omega (t) |$ ($\gamma < | \Delta \omega (t) |$) is always met, and hence the LE is negative (zero), i.e., stable (neutrally stable) at all times, as seen from Eq.~\eqref{eq:attr-instant-lyap}. However, the SIS (dashed) stays bounded only intermittently. In this example, the timescale that controls the alternation of those epochs is the period of the imposed modulation, $T_m = 2 \pi / \omega_m$, which could be arbitrarily long. Note that these epochs of growth are not the typical $2 \pi$ \textit{phase slips} of the fixed-frequency single oscillator case~\cite{pikovsky2003}, observed close to the synchronisation border. Here, the ``slip'' is a drift caused by the temporary neutral stability, and its relative importance depends on the length of that neutrally stable epoch, as seen from Fig.~\ref{fig:attr-intermittent}(b).

Now, the SIS is analysed further as shown in Fig.~\ref{fig:attr-intermittent}(b)-(d). Figure~\ref{fig:attr-intermittent}~(b) shows the adiabatic LE (dashed black)~\eqref{eq:attr-instant-lyap}. Epochs where it is negative (null) correspond to the $\psi_i$ growing (staying bounded). Moreover, Fig.~\ref{fig:attr-intermittent}~(b) shows the agreement between the instantaneous LE~\eqref{eq:insta_lambda} (grey) and the adiabatic LE~\eqref{eq:attr-instant-lyap}. This confirms \textit{a posteriori} the adequacy of the adiabatic approach. The intermittency can also be seen in the time-frequency representation of $\sin \theta_1 (t)$ for a trajectory of system~\eqref{eq:system}, as shown in Fig.~\ref{fig:attr-intermittent}(c). Here, stability epochs correspond to the frequency being entrained by the driver (one frequency mode), whereas neutral stability epochs correspond to the presence of two frequency modes plus harmonics. Finally, a negative maximum LE guarantees convergence of different initial conditions, as long as they are in the basin of attraction of the synchronous solution, even if the trajectory is only intermittently synchronised to the driving. This is illustrated in Fig.~\ref{fig:attr-intermittent}(d).

Note that all other non-maximal LEs of the spectrum are negative at all time, $\lambda^i(t)<0$, for $i=2,\ldots,N$, since the corresponding Laplacian eigenvalues are negative. The exponent $\lambda^1$ measures the stability against a homogeneous perturbation, whereas the rest of the spectrum corresponds to any heterogeneous perturbation. In other words, any synchronous solution will stay synchronous -- all oscillators with the same phase -- against any perturbation. During stable epochs, even the common phase of the synchronous state is stable. During neutrally stable epochs, however, the common phase of the synchronous solution can be pushed by a homogeneous perturbation, and change without the perturbation decaying or growing.

On average, over a time $T$, the maximum exponent is
\begin{equation}
\lambda_{max} = \tfrac{1}{T} \int_0^T \dd t \, \, \lambda_{max} (t) \le 0 .
\label{eq:attr-avg-lyap}
\end{equation}
Note that the LE in Eq.~\eqref{eq:attr-avg-lyap} is strictly zero only in region I where the system does not synchronise to the driving at any time. Indeed, in region III, the adiabatic LE alternates between zero and negative values, and is negative on average. Moreover, region I decreases in size as $k$ is increased, and so the remainder of parameter space, corresponding to stability $\lambda_{max}<0$, grows. In other words, by increasing the amplitude of the nonautonomous modulation, one makes the region of stability larger in parameter space. In this region of stability, different initial conditions converge to one unique trajectory. In Fig.~\ref{fig:fig-tongues-attr}, panels (a)-(c) show the enlargement of the negative LE region in parameter space as $k$ increases from $0.0$ to $0.4$, and $0.8$. Panel (d) combines  and shows those LE values for all three values of $k$, but for a single value of the forcing strength $\gamma = 2.5$. The region of stability is the union of regions II and III. Region II, where trajectories are always synchronised to the driving, decreases in size as $k$ is increased, but region III grows enough so that their union grows.  The phenomenon does not depend on the explicit form of $f(\omega_m t)$ and was also illustrated numerically, in the single driven oscillator case, for different aperiodic $f(\omega_m t)$, and $f(\omega_m t)$ defined over a finite time in \cite{lucas2018,newman2018}.

%\begin{figure}[t]
%	\centering
%	\includegraphics[width=\linewidth]{Fig2.pdf}
%	%{fig-tongues-attr_D_-0.5_w0_2.00_ks_[0.0, 0.4, 0.8]-slice}
%	\caption{Stability region increases with $k$, as shown by the LE from formula~\eqref{eq:attr-avg-lyap}, for $\gamma =2.5$, and different values of the frequency modulation amplitude $k$. The stable range of $\Delta \omega$, for which $\lambda_{max}<0$, increases as $k$ is increased. }
%	\label{fig:attr-slice}
%\end{figure}

\begin{figure*}[htb]
	\centering
	\includegraphics[width=\linewidth]{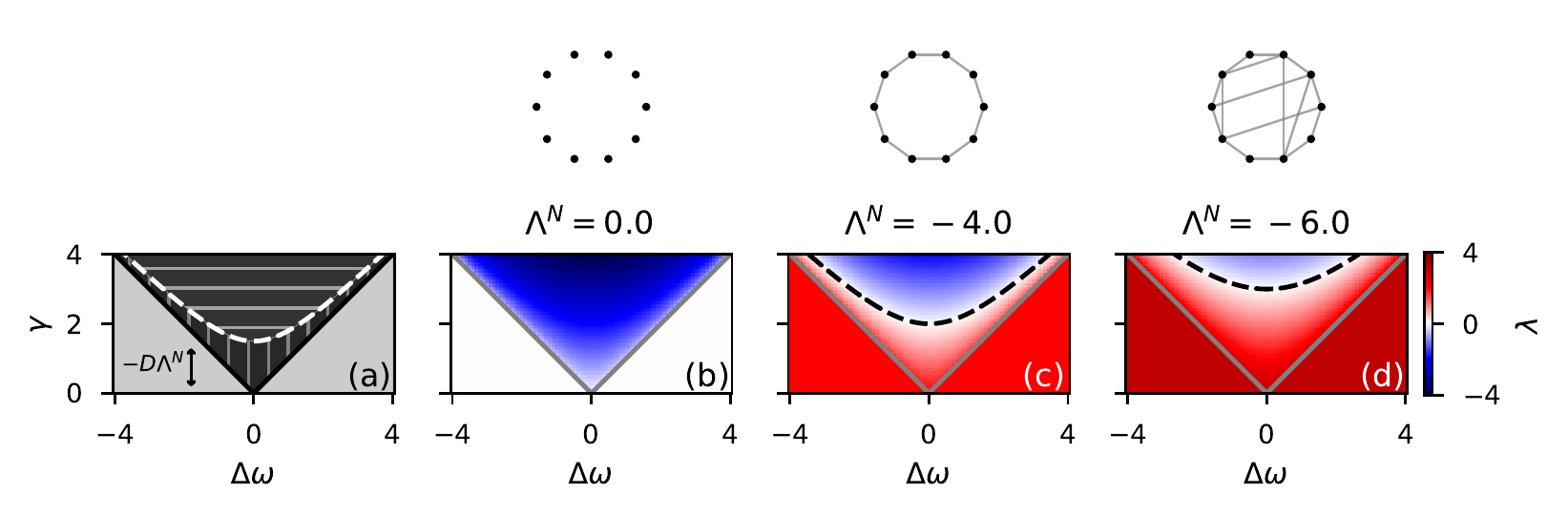}
	%{fig-tongues-repuls_D_0.5_w0_2.00_Lams_[0.0, -4.0, -6]-horizontal4.pdf}
	%{fig-tongues-repuls_D_0.5_w0_2.00_Lams_[0.0, -3.0, -6]-horizontal4_bis}
	\caption{Stability region shrinks as $\Lambda^N$ becomes more negative, i.e. as the largest degree of connectivity is increased, in the repulsive, $D>0$, autonomous, $k=0$, case. (a) Theoretical qualitative regions. Region I: no synchronisation (light grey). Region IIa: synchronisation unstable (dark grey vertically hatched). Region IIb: synchronisation stable (dark grey horizontally hatched). (b)-(d) Quantitative stability measured by the LE: the region of stability IIb (shades of blue) decreases with $\Lambda^N$, while the region of instability IIa (shades of red) increases. White values represent zero values of the LE. Above each panel, an example network that has the corresponding value of $\Lambda^N$ is shown. Parameters are $D=0.5$, $w_0=2$.}
	\label{fig:rep-tongues-a}
\end{figure*}

\begin{figure}[tb]
	\centering
	\includegraphics[width=\linewidth]{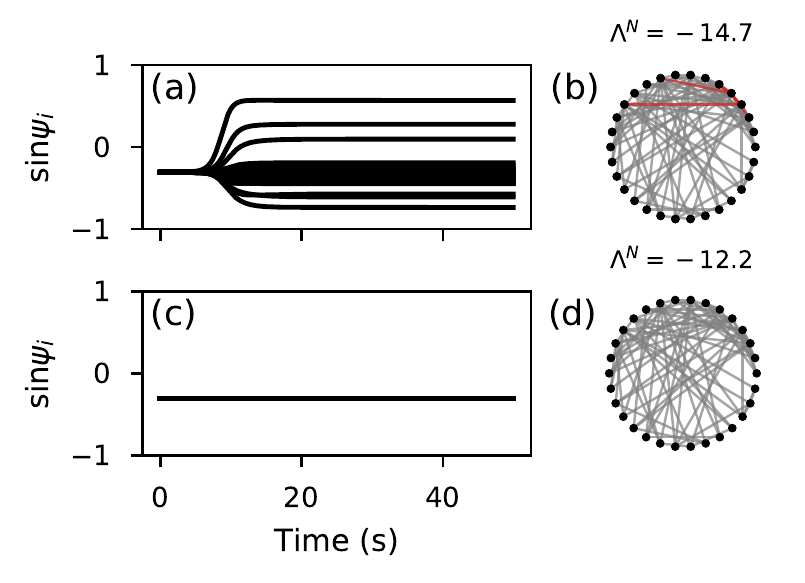}
	%{fig_cut_BA_N_30_D_0.5_gamma_6.50_dw_2.00_k_0.0_ic_sync}
	\caption{Control of the stability by cutting a few chosen links, in the repulsive case $D>0$. (a), (c) Trajectory of a synchronous synchronised initial condition of the phase difference $\psi_i$, in (b), (d) the corresponding network, respectively. In the initial network (b), the synchronous state is unstable (a). By cutting a few chosen links (red), in the new network (d), the synchronous state is made stable (c). The initial network is a Barabási-Albert with $N=30$, and 5 links are cut. Other parameters are $D=0.5$, $\gamma=6.5$, $\Delta \omega = 2$, $\omega_0=1$, $k=0$.}
	\label{fig:rep-cut}
\end{figure}

\begin{figure*}[htb]
	\centering
	\includegraphics[width=\linewidth]{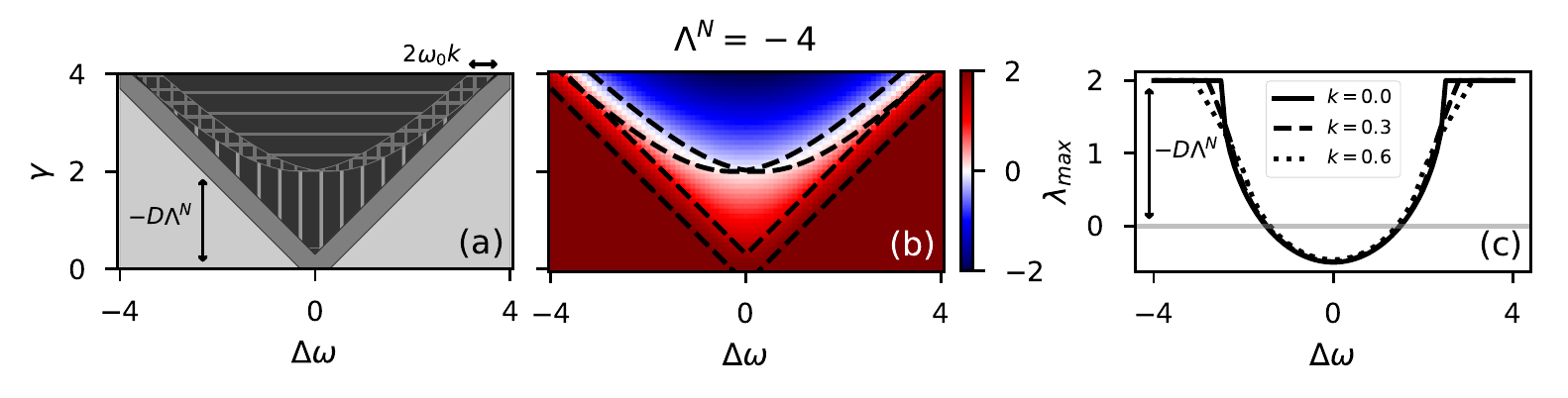}
	%{fig-tongues-repuls_D_0.5_w0_1.00_Lam_-4.00_ks_[0.0, 0.3, 0.6]-horizontal3}
	\caption{Stability regions, in the repulsive, $D>0$, nonautonomous, $k=3$, case. (a) Theoretical qualitative regions. Birth of region IIc (square hatch), where the SS is alternates between stability and instability. Other regions are I (light grey) and III (medium grey) [see Fig.~\ref{fig:fig-tongues-attr}], and IIa (vertical hatch) and IIb (horizontal hatch) [see Fig.~\ref{fig:rep-tongues-a}]. (b) Quantitative stability measured by the LE. White values represent zero values of the LE. (c) LE for a fixed $\gamma=2.5$ and different values of $k$. Other parameters are $D=0.5$, $w_0=1$.}
	\label{fig:rep-tongues-na}
\end{figure*}

\subsection{Repulsive case} \label{sec:repulsive}

In this case, $D>0$, all oscillators have a tendency towards pairwise asynchrony, when not driven. In the unforced network, the synchronous solution is always unstable, but if forced, the synchronous solution can be stable or unstable. The largest adiabatic LE, in this case $\lambda_{max}(t) = \lambda^N(t)$ corresponding to $\Lambda^N < 0$ in formula~\eqref{eq:instant-lyap}, is given by
\begin{equation}
\lambda_{max}(t) =
\left\{
\begin{array}{ll}
- D \Lambda^{N} - \sqrt{\gamma^2 - \Delta \omega(t)^2} & \text{for } t :  \gamma > | \Delta \omega (t) | , \\
- D \Lambda^{N} > 0 & \text{else} . \\
\end{array}
\right.
\label{eq:rep-instant-lyap}
\end{equation}
Firstly, when the driving strength is not large enough, i.e., $\gamma < |\Delta \omega (t)|$, synchronous solutions are unstable -- i.e., always for the SNS and during the non-synchronised epochs for the SIS. This is in contrast with the attractive case, which exhibited neutral stability under the same condition. Secondly, when $\gamma > |\Delta \omega (t)|$, two effects compete: the network couplings push oscillators away from each other (positive first term) whereas the external driving brings them back towards a synchronous state (negative second term). This is again in contrast with the attractive case, which only exhibits stability under the same condition. Those two effects exactly compensate each other if
\begin{equation}
\gamma = \sqrt{(D \Lambda^N)^2  + \Delta \omega^2(t)} ,
\label{eq:zero}
\end{equation}
which has a minimum when $\Delta \omega = 0$ at $\gamma = -D \Lambda^N$. This is shown for the autonomous case $k=0$ in Fig.~\ref{fig:rep-tongues-a}(a) (dashed white).

In the autonomous case, $k=0$, the frequency mismatch is constant in time. Hence, condition~\eqref{eq:zero} divides region II into two subregions based on stability: regions IIa (IIb) where the SS is unstable (stable) corresponding to $\gamma$ values smaller (larger) than that of condition~\eqref{eq:zero}. The regions are shown in parameter space in Fig.~\ref{fig:rep-tongues-a}(a), and are confirmed by the computation of the $\lambda_{max}$, as shown Fig.~\ref{fig:rep-tongues-a}(b)-(d) for different values of $\Lambda^N$. As the largest-magnitude eigenvalue $\Lambda^N$ of the Laplacian, or alternatively the network coupling strength $D$, is increased, the region of negative $\lambda_{max}$ decreases in size.

This observation can be used as a viable control strategy. Indeed, note that the eigenvalues of the Laplacian, and in particular $\Lambda^N$, are determined by the topology of the network considered. Moreover, the inequality $| \Lambda^N | \le 2 K_{max}$ holds, where $K_{max}$ is the number of connections of the most connected node~\cite{almendral2007}. In other words, the stability of the dynamics is directly determined by the topology, and in particular by the connectivity of the most connected node. Here, in the repulsive case, more connections between nodes, and in particular to the most connected one, means less stability.  So then, one can optimally decrease the absolute value of $\Lambda^N$ -- or equivalently increase the region of stability --  by cutting the edges of the most central node, which in turn amounts to reducing $K_{max}$. This is illustrated in Fig.~\ref{fig:rep-cut} where the SS is stabilised by cutting only 5 chosen links (red) out of 81 ($\approx 6 \%$). The original network is a Barabási-Albert one [see Fig.~\ref{fig:rep-cut}(b)] \footnote{The network topology was created with the Python function \texttt{barabasi\_albert\_graph(N, m)} from the \texttt{NetworkX} package~\cite{scipy_proceedings11}, with $N=30$ the total number of nodes, and $m=3$ the number of nodes added.}, for which this strategy is most effective, since only a few nodes have very high connectivity.

As a by-product of the analysis, in region IIa, we numerically observed partially-locked states~\cite{ko2008}, or chimera-like states~\cite{kuramoto2002,panaggio2015}, where most of the oscillators are phase-locked to the driving in a quasi-synchronous cluster, while the rest of the oscillators drift independently. An example video of such dynamics is provided in the Supplementary Material.

In general, in the nonautonomous case, $k\neq0$, an additional type of intermittent synchronisation can be exhibited, which here alternates between stability and instability. That is, the adiabatic LE for the synchronous solutions alternates between negative and positive values. This happens in a new region IIc which, as $k$ is made progressively larger than $0$, appears at the border between regions IIa and IIb and subsequently grows, as illustrated in Fig.~\ref{fig:rep-tongues-na}(a) in squared grey hatch. Region IIc can be defined as all pairs $(\gamma, \Delta \omega)$ such that $\gamma$ is greater than the value of condition~\eqref{eq:zero} at certain times and smaller at others. Equivalently, it is $\{(\gamma, \Delta \omega) : \exists \, t : \gamma = \sqrt{(D \Lambda^N)^2  + \Delta \omega^2(t)} \}$. The region is constant in time, and the explicit form of its boundaries is uninformative and is hence omitted here. In region IIc, the long-term LE can be either positive or negative. For small enough values of the coupling strength $D$, the region of negative LE increases with $k$, just as in the attractive case of Sec.~\ref{sec:attractive}. This effect can be achieved for relatively small values of the coupling strength $D$.

\begin{figure}[t]
	\centering
	\includegraphics[width=\linewidth]{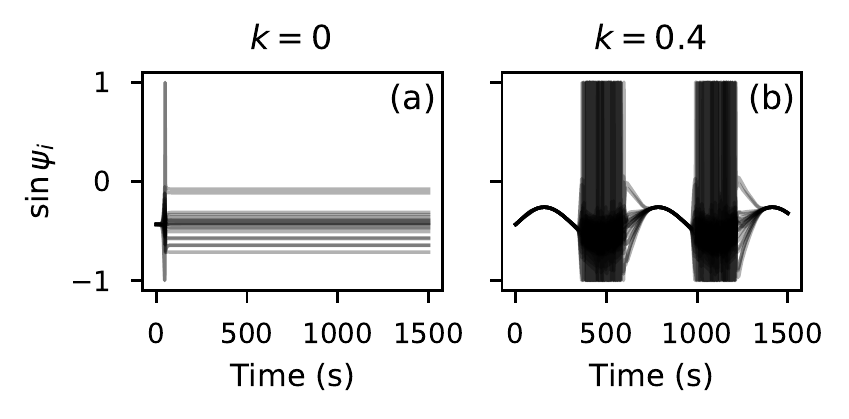}
	%{fig-intermittent_net_psi_random_N_20_D_0.5_gamma_4.60_dw_2.00_k_0.4_ic_sync}
	\caption{Control strategy: fully synchronous state enforced intermittently by nonautonomous driving. Trajectory of a synchronous initial condition for $N=20$ oscillators for (a) constant and (b) modulated driving frequency. (a) Synchronous solution is unstable and yields patterns. (b) Time-variability of the driving forces state to be synchronous intermittently. Other parameters are $D=0.5$, $\gamma=4.6$, $\Delta \omega = 2$, $\omega_0=2$.  Network used is random with $p=0.2$ and $\Lambda^N = -8.68$.}
	\label{fig:repuls-alternation}
\end{figure}

However, while alternation between neutral stability and stability guaranteed overall stability and convergence in the previous section, alternation between stability and instability does not. Here, in region IIc, for a synchronous initial condition, a negative LE associated to its trajectory does not guarantee the state will stay synchronous and converge to the SS solution, as shown in Fig.~\ref{fig:repuls-alternation}(b). Indeed, during epochs of instability, any perturbation can push the state far away from its original synchronous solution exponentially fast. 
However, if no other attractor exists during the epochs of stability, then the system will return to the synchronised synchronous state when the next epoch of stability occurs. Thus it alternates between the two regimes, as shown numerically in Fig.~\ref{fig:repuls-alternation}(b). This effect is purely due to the time-variability of the driving, controlled by $k$. Moreover, there is subregion in parameter space which is unstable (region IIa) when $k=0$, but turns intermittently stable (region IIc) for $k>0$. This can be used as a control strategy as illustrated in Fig.~\ref{fig:repuls-alternation}: one forces a completely incoherent state [panel (a)] to be synchronous and synchronised indefinitely often and long, by allowing time-variability in the driving frequency. Thus, time-variability can also be used to counter-balance the desynchronising effect which is produced by a highly connected network, as shown in Fig.~\ref{fig:rep-cut}. Finally, this strategy can be useful when the requirement for synchronicity are not so stringent and the system under inspection does not need to be synchronous at all times.

\section{Summary and conclusions} \label{sec:conclusion}

In this work, we studied the effect of driving an arbitrary network of identical phase oscillators with an external time-varying-frequency oscillator. This is in contrast with previous studies of forced networks, which do not consider time-variability, a crucial ingredient in real-world systems. Stability -- both short- and long-term -- of the synchronous solutions was assessed by linear stability analysis, and results were confirmed via numerical simulations. The system studied is nonautonomous and hard to treat in general, hence we assumed slow time-variability. Two cases were treated: attractive and repulsive couplings, where oscillators tend to be in phase and out of phase, respectively.

In the attractive case, we showed that increasing the amplitude $k$ of the frequency modulation enlarged the stability region in parameter space, as defined by a negative Lyapunov exponent. In other words, a more variable driving makes the stability of synchronous behaviour more robust against parameter changes. This is the direct generalisation to networks of a result that was obtained for a single driven oscillator. An additional region of intermittency between stability and neutral stability appears, as a result of the time-variability in the frequency.

In the repulsive case, we first demonstrated a control strategy, where the synchronous solution can be made stable by cutting a few chosen links in the network. Moreover, we established the phase diagram of the synchronous solutions. We then showed that one could counter-balance the desynchronising effect of connectivity by allowing time-variability in the driving frequency. Indeed, we showed that the time-variability of the frequency can drive the system back to a synchronous state intermittently, where it was asynchronous with a fixed-frequency driving. This observation can serve as an alternative control protocol. Finally, as a by-product of the analysis, we numerically observed chimera-like states.

Dynamics of the phase and frequency is illustrated in detail in the different cases. Such classification can potentially be of use to experimentalists who can only measure phase and frequency, and may have only limited or no knowledge of an external driving. Moreover, we illustrated diverse and simple control strategies to enhance synchronisability that could explain how living systems maintain stability in a changing environment, and could also be implemented directly.

Finally, many of the interesting and physically relevant features observed in this system were either happening over finite time, or explained by the finite-time analysis of dynamics. Consequently, a solely asymptotic analysis would have missed much of the dynamical intricacies at hand. From this, we conclude that nonautonomous networks of coupled oscillators, such as the present one, exhibit features reminiscent of those observed in living systems, and that a finite-time approach is crucial to their understanding.

%\vspace{1.5em}

\section*{Acknowledgments}

We thank Julian Newman and Yevhen Suprunenko for very useful discussions and comments on the manuscript. This work has been funded by the EU's Horizon 2020 research and innovation programme under the Marie Sk\l{}odowska-Curie grant agreement No 642563 and the EPSRC grant EP/M006298/1, UK.

%\clearpage
%=========================
%\bibliographystyle{unsrt}
%\bibliographystyle{aipauth4-1}
\bibliographystyle{apsrev4-1}

%\nocite{*}
\bibliography{bib}

%merlin.mbs apsrev4-1.bst 2010-07-25 4.21a (PWD, AO, DPC) hacked
%Control: key (0)
%Control: author (72) initials jnrlst
%Control: editor formatted (1) identically to author
%Control: production of article title (-1) disabled
%Control: page (0) single
%Control: year (1) truncated
%Control: production of eprint (0) enabled
\begin{thebibliography}{46}%
\makeatletter
\providecommand \@ifxundefined [1]{%
 \@ifx{#1\undefined}
}%
\providecommand \@ifnum [1]{%
 \ifnum #1\expandafter \@firstoftwo
 \else \expandafter \@secondoftwo
 \fi
}%
\providecommand \@ifx [1]{%
 \ifx #1\expandafter \@firstoftwo
 \else \expandafter \@secondoftwo
 \fi
}%
\providecommand \natexlab [1]{#1}%
\providecommand \enquote  [1]{``#1''}%
\providecommand \bibnamefont  [1]{#1}%
\providecommand \bibfnamefont [1]{#1}%
\providecommand \citenamefont [1]{#1}%
\providecommand \href@noop [0]{\@secondoftwo}%
\providecommand \href [0]{\begingroup \@sanitize@url \@href}%
\providecommand \@href[1]{\@@startlink{#1}\@@href}%
\providecommand \@@href[1]{\endgroup#1\@@endlink}%
\providecommand \@sanitize@url [0]{\catcode `\\12\catcode `\$12\catcode
  `\&12\catcode `\#12\catcode `\^12\catcode `\_12\catcode `\%12\relax}%
\providecommand \@@startlink[1]{}%
\providecommand \@@endlink[0]{}%
\providecommand \url  [0]{\begingroup\@sanitize@url \@url }%
\providecommand \@url [1]{\endgroup\@href {#1}{\urlprefix }}%
\providecommand \urlprefix  [0]{URL }%
\providecommand \Eprint [0]{\href }%
\providecommand \doibase [0]{http://dx.doi.org/}%
\providecommand \selectlanguage [0]{\@gobble}%
\providecommand \bibinfo  [0]{\@secondoftwo}%
\providecommand \bibfield  [0]{\@secondoftwo}%
\providecommand \translation [1]{[#1]}%
\providecommand \BibitemOpen [0]{}%
\providecommand \bibitemStop [0]{}%
\providecommand \bibitemNoStop [0]{.\EOS\space}%
\providecommand \EOS [0]{\spacefactor3000\relax}%
\providecommand \BibitemShut  [1]{\csname bibitem#1\endcsname}%
\let\auto@bib@innerbib\@empty
%</preamble>
\bibitem [{\citenamefont {Haken}(1975)}]{Haken:75}%
  \BibitemOpen
  \bibfield  {author} {\bibinfo {author} {\bibfnamefont {H.}~\bibnamefont
  {Haken}},\ }\href {\doibase 10.1103/RevModPhys.47.67} {\bibfield  {journal}
  {\bibinfo  {journal} {Rev.\ Mod. Phys.}\ }\textbf {\bibinfo {volume} {47}},\
  \bibinfo {pages} {67} (\bibinfo {year} {1975})}\BibitemShut {NoStop}%
\bibitem [{\citenamefont {Nicolis}\ and\ \citenamefont
  {Prigogine}(1977)}]{nicolis1977}%
  \BibitemOpen
  \bibfield  {author} {\bibinfo {author} {\bibfnamefont {G.}~\bibnamefont
  {Nicolis}}\ and\ \bibinfo {author} {\bibfnamefont {I.}~\bibnamefont
  {Prigogine}},\ }\href@noop {} {\emph {\bibinfo {title} {Self-organization in
  nonequilibrium systems}}}\ (\bibinfo  {publisher} {Wiley, New York},\
  \bibinfo {year} {1977})\BibitemShut {NoStop}%
\bibitem [{\citenamefont {Goldbeter}(1997)}]{goldbeter1997}%
  \BibitemOpen
  \bibfield  {author} {\bibinfo {author} {\bibfnamefont {A.}~\bibnamefont
  {Goldbeter}},\ }\href@noop {} {\emph {\bibinfo {title} {{Biochemical
  Oscillations and Cellular Rhythms}}}}\ (\bibinfo  {publisher} {Cambridge
  University Press},\ \bibinfo {address} {Cambridge},\ \bibinfo {year}
  {1997})\BibitemShut {NoStop}%
\bibitem [{\citenamefont {Pikovsky}\ \emph {et~al.}(2003)\citenamefont
  {Pikovsky}, \citenamefont {Rosenblum},\ and\ \citenamefont
  {Kurths}}]{pikovsky2003}%
  \BibitemOpen
  \bibfield  {author} {\bibinfo {author} {\bibfnamefont {A.}~\bibnamefont
  {Pikovsky}}, \bibinfo {author} {\bibfnamefont {M.}~\bibnamefont {Rosenblum}},
  \ and\ \bibinfo {author} {\bibfnamefont {J.}~\bibnamefont {Kurths}},\
  }\href@noop {} {\emph {\bibinfo {title} {{Synchronization: A Universal
  Concept in Nonlinear Sciences}}}},\ Vol.~\bibinfo {volume} {12}\ (\bibinfo
  {publisher} {Cambridge University Press},\ \bibinfo {address} {Cambridge},\
  \bibinfo {year} {2003})\BibitemShut {NoStop}%
\bibitem [{\citenamefont {Strogatz}(2004)}]{strogatz2004}%
  \BibitemOpen
  \bibfield  {author} {\bibinfo {author} {\bibfnamefont {S.~H.}\ \bibnamefont
  {Strogatz}},\ }\href@noop {} {\emph {\bibinfo {title} {Sync: The emerging
  science of spontaneous order}}}\ (\bibinfo  {publisher} {Penguin UK},\
  \bibinfo {year} {2004})\BibitemShut {NoStop}%
\bibitem [{\citenamefont {Iatsenko}\ \emph {et~al.}(2013)\citenamefont
  {Iatsenko}, \citenamefont {Bernjak}, \citenamefont {Stankovski},
  \citenamefont {Shiogai}, \citenamefont {Owen-Lynch}, \citenamefont
  {Clarkson}, \citenamefont {McClintock},\ and\ \citenamefont
  {Stefanovska}}]{Iatsenko:13a}%
  \BibitemOpen
  \bibfield  {author} {\bibinfo {author} {\bibfnamefont {D.}~\bibnamefont
  {Iatsenko}}, \bibinfo {author} {\bibfnamefont {A.}~\bibnamefont {Bernjak}},
  \bibinfo {author} {\bibfnamefont {T.}~\bibnamefont {Stankovski}}, \bibinfo
  {author} {\bibfnamefont {Y.}~\bibnamefont {Shiogai}}, \bibinfo {author}
  {\bibfnamefont {P.~J.}\ \bibnamefont {Owen-Lynch}}, \bibinfo {author}
  {\bibfnamefont {P.~B.~M.}\ \bibnamefont {Clarkson}}, \bibinfo {author}
  {\bibfnamefont {P.~V.~E.}\ \bibnamefont {McClintock}}, \ and\ \bibinfo
  {author} {\bibfnamefont {A.}~\bibnamefont {Stefanovska}},\ }\href {\doibase
  10.1098/rsta.2011.0622} {\bibfield  {journal} {\bibinfo  {journal} {Phil.
  Trans. R. Soc. Lond. A}\ }\textbf {\bibinfo {volume} {371}},\ \bibinfo
  {pages} {20110622} (\bibinfo {year} {2013})}\BibitemShut {NoStop}%
\bibitem [{\citenamefont {Stankovski}\ \emph {et~al.}(2016)\citenamefont
  {Stankovski}, \citenamefont {Petkoski}, \citenamefont {Raeder}, \citenamefont
  {Smith}, \citenamefont {McClintock},\ and\ \citenamefont
  {Stefanovska}}]{Stankovski:16}%
  \BibitemOpen
  \bibfield  {author} {\bibinfo {author} {\bibfnamefont {T.}~\bibnamefont
  {Stankovski}}, \bibinfo {author} {\bibfnamefont {S.}~\bibnamefont
  {Petkoski}}, \bibinfo {author} {\bibfnamefont {J.}~\bibnamefont {Raeder}},
  \bibinfo {author} {\bibfnamefont {A.~F.}\ \bibnamefont {Smith}}, \bibinfo
  {author} {\bibfnamefont {P.~V.~E.}\ \bibnamefont {McClintock}}, \ and\
  \bibinfo {author} {\bibfnamefont {A.}~\bibnamefont {Stefanovska}},\ }\href
  {\doibase 10.1098/rsta.2015.0186} {\bibfield  {journal} {\bibinfo  {journal}
  {Phil. Trans. R. Soc. A}\ }\textbf {\bibinfo {volume} {374}},\ \bibinfo
  {pages} {20150186} (\bibinfo {year} {2016})}\BibitemShut {NoStop}%
\bibitem [{\citenamefont {Mormann}\ \emph {et~al.}(2000)\citenamefont
  {Mormann}, \citenamefont {Lehnertz}, \citenamefont {David},\ and\
  \citenamefont {Elger}}]{mormann2000}%
  \BibitemOpen
  \bibfield  {author} {\bibinfo {author} {\bibfnamefont {F.}~\bibnamefont
  {Mormann}}, \bibinfo {author} {\bibfnamefont {K.}~\bibnamefont {Lehnertz}},
  \bibinfo {author} {\bibfnamefont {P.}~\bibnamefont {David}}, \ and\ \bibinfo
  {author} {\bibfnamefont {C.~E.}\ \bibnamefont {Elger}},\ }\href {\doibase
  10.1016/S0167-2789(00)00087-7} {\bibfield  {journal} {\bibinfo  {journal}
  {Phys. D}\ }\textbf {\bibinfo {volume} {144}},\ \bibinfo {pages} {358}
  (\bibinfo {year} {2000})}\BibitemShut {NoStop}%
\bibitem [{\citenamefont {Andrzejak}\ \emph {et~al.}(2016)\citenamefont
  {Andrzejak}, \citenamefont {Rummel}, \citenamefont {Mormann},\ and\
  \citenamefont {Schindler}}]{andrzejak2016}%
  \BibitemOpen
  \bibfield  {author} {\bibinfo {author} {\bibfnamefont {R.~G.}\ \bibnamefont
  {Andrzejak}}, \bibinfo {author} {\bibfnamefont {C.}~\bibnamefont {Rummel}},
  \bibinfo {author} {\bibfnamefont {F.}~\bibnamefont {Mormann}}, \ and\
  \bibinfo {author} {\bibfnamefont {K.}~\bibnamefont {Schindler}},\ }\href
  {\doibase 10.1038/srep23000} {\bibfield  {journal} {\bibinfo  {journal} {Sci.
  Rep.}\ }\textbf {\bibinfo {volume} {6}},\ \bibinfo {pages} {23000} (\bibinfo
  {year} {2016})}\BibitemShut {NoStop}%
\bibitem [{\citenamefont {Stankovski}\ \emph {et~al.}(2017)\citenamefont
  {Stankovski}, \citenamefont {Pereira}, \citenamefont {McClintock},\ and\
  \citenamefont {Stefanovska}}]{Stankovski2017}%
  \BibitemOpen
  \bibfield  {author} {\bibinfo {author} {\bibfnamefont {T.}~\bibnamefont
  {Stankovski}}, \bibinfo {author} {\bibfnamefont {T.}~\bibnamefont {Pereira}},
  \bibinfo {author} {\bibfnamefont {P.~V.~E.}\ \bibnamefont {McClintock}}, \
  and\ \bibinfo {author} {\bibfnamefont {A.}~\bibnamefont {Stefanovska}},\
  }\href {\doibase 10.1103/RevModPhys.89.045001} {\bibfield  {journal}
  {\bibinfo  {journal} {Rev. Mod. Phys.}\ }\textbf {\bibinfo {volume} {89}},\
  \bibinfo {pages} {045001} (\bibinfo {year} {2017})}\BibitemShut {NoStop}%
\bibitem [{\citenamefont {Kuramoto}(1984)}]{kuramoto1984}%
  \BibitemOpen
  \bibfield  {author} {\bibinfo {author} {\bibfnamefont {Y.}~\bibnamefont
  {Kuramoto}},\ }\href@noop {} {\emph {\bibinfo {title} {Chemical oscillations,
  waves, and turbulence}}}\ (\bibinfo  {publisher} {Springer-Verlag},\ \bibinfo
  {address} {Tokyo},\ \bibinfo {year} {1984})\BibitemShut {NoStop}%
\bibitem [{\citenamefont {Arenas}\ \emph {et~al.}(2008)\citenamefont {Arenas},
  \citenamefont {D{\'{i}}az-Guilera}, \citenamefont {Kurths}, \citenamefont
  {Moreno},\ and\ \citenamefont {Zhou}}]{arenas2008}%
  \BibitemOpen
  \bibfield  {author} {\bibinfo {author} {\bibfnamefont {A.}~\bibnamefont
  {Arenas}}, \bibinfo {author} {\bibfnamefont {A.}~\bibnamefont
  {D{\'{i}}az-Guilera}}, \bibinfo {author} {\bibfnamefont {J.}~\bibnamefont
  {Kurths}}, \bibinfo {author} {\bibfnamefont {Y.}~\bibnamefont {Moreno}}, \
  and\ \bibinfo {author} {\bibfnamefont {C.}~\bibnamefont {Zhou}},\ }\href
  {\doibase 10.1016/j.physrep.2008.09.002} {\bibfield  {journal} {\bibinfo
  {journal} {Phys. Rep.}\ }\textbf {\bibinfo {volume} {469}},\ \bibinfo {pages}
  {93} (\bibinfo {year} {2008})}\BibitemShut {NoStop}%
\bibitem [{\citenamefont {D{\"{o}}rfler}\ and\ \citenamefont
  {Bullo}(2014)}]{dorfler2014}%
  \BibitemOpen
  \bibfield  {author} {\bibinfo {author} {\bibfnamefont {F.}~\bibnamefont
  {D{\"{o}}rfler}}\ and\ \bibinfo {author} {\bibfnamefont {F.}~\bibnamefont
  {Bullo}},\ }\href {\doibase 10.1016/j.automatica.2014.04.012} {\bibfield
  {journal} {\bibinfo  {journal} {Automatica}\ }\textbf {\bibinfo {volume}
  {50}},\ \bibinfo {pages} {1539} (\bibinfo {year} {2014})}\BibitemShut
  {NoStop}%
\bibitem [{\citenamefont {Rodrigues}\ \emph {et~al.}(2016)\citenamefont
  {Rodrigues}, \citenamefont {Peron}, \citenamefont {Ji},\ and\ \citenamefont
  {Kurths}}]{rodrigues2016}%
  \BibitemOpen
  \bibfield  {author} {\bibinfo {author} {\bibfnamefont {F.~A.}\ \bibnamefont
  {Rodrigues}}, \bibinfo {author} {\bibfnamefont {T.~K. D.~M.}\ \bibnamefont
  {Peron}}, \bibinfo {author} {\bibfnamefont {P.}~\bibnamefont {Ji}}, \ and\
  \bibinfo {author} {\bibfnamefont {J.}~\bibnamefont {Kurths}},\ }\href
  {\doibase 10.1016/j.physrep.2015.10.008} {\bibfield  {journal} {\bibinfo
  {journal} {Phys. Rep.}\ }\textbf {\bibinfo {volume} {610}},\ \bibinfo {pages}
  {1} (\bibinfo {year} {2016})}\BibitemShut {NoStop}%
\bibitem [{\citenamefont {Glass}(2001)}]{Glass2001}%
  \BibitemOpen
  \bibfield  {author} {\bibinfo {author} {\bibfnamefont {L.}~\bibnamefont
  {Glass}},\ }\href {\doibase 10.1038/35065745} {\bibfield  {journal} {\bibinfo
   {journal} {Nature}\ }\textbf {\bibinfo {volume} {410}},\ \bibinfo {pages}
  {277} (\bibinfo {year} {2001})}\BibitemShut {NoStop}%
\bibitem [{\citenamefont {Leloup}\ \emph {et~al.}(1999)\citenamefont {Leloup},
  \citenamefont {Gonze},\ and\ \citenamefont {Goldbeter}}]{leloup1999}%
  \BibitemOpen
  \bibfield  {author} {\bibinfo {author} {\bibfnamefont {J.-C.}\ \bibnamefont
  {Leloup}}, \bibinfo {author} {\bibfnamefont {D.}~\bibnamefont {Gonze}}, \
  and\ \bibinfo {author} {\bibfnamefont {A.}~\bibnamefont {Goldbeter}},\ }\href
  {\doibase 10.1177/074873099129000948} {\bibfield  {journal} {\bibinfo
  {journal} {J. Biol. Rhythms}\ }\textbf {\bibinfo {volume} {14}},\ \bibinfo
  {pages} {433} (\bibinfo {year} {1999})}\BibitemShut {NoStop}%
\bibitem [{\citenamefont {{Bra{\v{c}}i{\v{c}} Lotri{\v{c}}}}\ and\
  \citenamefont {Stefanovska}(2000)}]{Stefanovska2000}%
  \BibitemOpen
  \bibfield  {author} {\bibinfo {author} {\bibfnamefont {M.}~\bibnamefont
  {{Bra{\v{c}}i{\v{c}} Lotri{\v{c}}}}}\ and\ \bibinfo {author} {\bibfnamefont
  {A.}~\bibnamefont {Stefanovska}},\ }\href
  {http://dx.doi.org/10.1016/S0378-4371(00)00204-1} {\bibfield  {journal}
  {\bibinfo  {journal} {Phys. A}\ }\textbf {\bibinfo {volume} {283}},\ \bibinfo
  {pages} {451} (\bibinfo {year} {2000})}\BibitemShut {NoStop}%
\bibitem [{\citenamefont {Stefanovska}\ \emph {et~al.}(2001)\citenamefont
  {Stefanovska}, \citenamefont {{Bra{\v{c}}i{\v{c}} Lotri\v{c}}}, \citenamefont
  {Strle},\ and\ \citenamefont {Haken}}]{Stefanovska:01a}%
  \BibitemOpen
  \bibfield  {author} {\bibinfo {author} {\bibfnamefont {A.}~\bibnamefont
  {Stefanovska}}, \bibinfo {author} {\bibfnamefont {M.}~\bibnamefont
  {{Bra{\v{c}}i{\v{c}} Lotri\v{c}}}}, \bibinfo {author} {\bibfnamefont
  {S.}~\bibnamefont {Strle}}, \ and\ \bibinfo {author} {\bibfnamefont
  {H.}~\bibnamefont {Haken}},\ }\href {\doibase 10.1088/0967-3334/22/3/311}
  {\bibfield  {journal} {\bibinfo  {journal} {Physiol. Meas.}\ }\textbf
  {\bibinfo {volume} {22}},\ \bibinfo {pages} {535} (\bibinfo {year}
  {2001})}\BibitemShut {NoStop}%
\bibitem [{\citenamefont {Suprunenko}\ \emph {et~al.}(2013)\citenamefont
  {Suprunenko}, \citenamefont {Clemson},\ and\ \citenamefont
  {Stefanovska}}]{suprunenko2013}%
  \BibitemOpen
  \bibfield  {author} {\bibinfo {author} {\bibfnamefont {Y.~F.}\ \bibnamefont
  {Suprunenko}}, \bibinfo {author} {\bibfnamefont {P.~T.}\ \bibnamefont
  {Clemson}}, \ and\ \bibinfo {author} {\bibfnamefont {A.}~\bibnamefont
  {Stefanovska}},\ }\href {\doibase 10.1103/PhysRevLett.111.024101} {\bibfield
  {journal} {\bibinfo  {journal} {Phys. Rev. Lett.}\ }\textbf {\bibinfo
  {volume} {111}},\ \bibinfo {pages} {024101} (\bibinfo {year}
  {2013})}\BibitemShut {NoStop}%
\bibitem [{\citenamefont {Lancaster}\ \emph {et~al.}(2016)\citenamefont
  {Lancaster}, \citenamefont {Suprunenko}, \citenamefont {Jenkins},\ and\
  \citenamefont {Stefanovska}}]{lancaster2016}%
  \BibitemOpen
  \bibfield  {author} {\bibinfo {author} {\bibfnamefont {G.}~\bibnamefont
  {Lancaster}}, \bibinfo {author} {\bibfnamefont {Y.~F.}\ \bibnamefont
  {Suprunenko}}, \bibinfo {author} {\bibfnamefont {K.}~\bibnamefont {Jenkins}},
  \ and\ \bibinfo {author} {\bibfnamefont {A.}~\bibnamefont {Stefanovska}},\
  }\href {\doibase 10.1038/srep29584} {\bibfield  {journal} {\bibinfo
  {journal} {Sci. Rep.}\ }\textbf {\bibinfo {volume} {6}},\ \bibinfo {pages}
  {29584} (\bibinfo {year} {2016})}\BibitemShut {NoStop}%
\bibitem [{\citenamefont {Flourakis}\ \emph {et~al.}(2015)\citenamefont
  {Flourakis}, \citenamefont {Kula-Eversole}, \citenamefont {Hutchison},
  \citenamefont {Han}, \citenamefont {Aranda}, \citenamefont {Moose},
  \citenamefont {White}, \citenamefont {Dinner}, \citenamefont {Lear},
  \citenamefont {Ren}, \citenamefont {Diekman}, \citenamefont {Raman},\ and\
  \citenamefont {Allada}}]{flourakis2015}%
  \BibitemOpen
  \bibfield  {author} {\bibinfo {author} {\bibfnamefont {M.}~\bibnamefont
  {Flourakis}}, \bibinfo {author} {\bibfnamefont {E.}~\bibnamefont
  {Kula-Eversole}}, \bibinfo {author} {\bibfnamefont {A.~L.}\ \bibnamefont
  {Hutchison}}, \bibinfo {author} {\bibfnamefont {T.~H.}\ \bibnamefont {Han}},
  \bibinfo {author} {\bibfnamefont {K.}~\bibnamefont {Aranda}}, \bibinfo
  {author} {\bibfnamefont {D.~L.}\ \bibnamefont {Moose}}, \bibinfo {author}
  {\bibfnamefont {K.~P.}\ \bibnamefont {White}}, \bibinfo {author}
  {\bibfnamefont {A.~R.}\ \bibnamefont {Dinner}}, \bibinfo {author}
  {\bibfnamefont {B.~C.}\ \bibnamefont {Lear}}, \bibinfo {author}
  {\bibfnamefont {D.}~\bibnamefont {Ren}}, \bibinfo {author} {\bibfnamefont
  {C.~O.}\ \bibnamefont {Diekman}}, \bibinfo {author} {\bibfnamefont {I.~M.}\
  \bibnamefont {Raman}}, \ and\ \bibinfo {author} {\bibfnamefont
  {R.}~\bibnamefont {Allada}},\ }\href {\doibase 10.1016/j.cell.2015.07.036}
  {\bibfield  {journal} {\bibinfo  {journal} {Cell}\ }\textbf {\bibinfo
  {volume} {162}},\ \bibinfo {pages} {836} (\bibinfo {year}
  {2015})}\BibitemShut {NoStop}%
\bibitem [{\citenamefont {Pfeuty}\ \emph {et~al.}(2003)\citenamefont {Pfeuty},
  \citenamefont {Mato}, \citenamefont {Golomb},\ and\ \citenamefont
  {Hansel}}]{pfeuty2003}%
  \BibitemOpen
  \bibfield  {author} {\bibinfo {author} {\bibfnamefont {B.}~\bibnamefont
  {Pfeuty}}, \bibinfo {author} {\bibfnamefont {G.}~\bibnamefont {Mato}},
  \bibinfo {author} {\bibfnamefont {D.}~\bibnamefont {Golomb}}, \ and\ \bibinfo
  {author} {\bibfnamefont {D.}~\bibnamefont {Hansel}},\ }\href {\doibase
  10.1523/JNEUROSCI.23-15-06280.2003} {\bibfield  {journal} {\bibinfo
  {journal} {J. Neurosci.}\ }\textbf {\bibinfo {volume} {23}},\ \bibinfo
  {pages} {6280} (\bibinfo {year} {2003})}\BibitemShut {NoStop}%
\bibitem [{\citenamefont {Ushio}\ \emph {et~al.}(2018)\citenamefont {Ushio},
  \citenamefont {Hsieh}, \citenamefont {Masuda}, \citenamefont {Deyle},
  \citenamefont {Ye}, \citenamefont {Chang}, \citenamefont {Sugihara},\ and\
  \citenamefont {Kondoh}}]{ushio2018}%
  \BibitemOpen
  \bibfield  {author} {\bibinfo {author} {\bibfnamefont {M.}~\bibnamefont
  {Ushio}}, \bibinfo {author} {\bibfnamefont {C.-H.}\ \bibnamefont {Hsieh}},
  \bibinfo {author} {\bibfnamefont {R.}~\bibnamefont {Masuda}}, \bibinfo
  {author} {\bibfnamefont {E.~R.}\ \bibnamefont {Deyle}}, \bibinfo {author}
  {\bibfnamefont {H.}~\bibnamefont {Ye}}, \bibinfo {author} {\bibfnamefont
  {C.-W.}\ \bibnamefont {Chang}}, \bibinfo {author} {\bibfnamefont
  {G.}~\bibnamefont {Sugihara}}, \ and\ \bibinfo {author} {\bibfnamefont
  {M.}~\bibnamefont {Kondoh}},\ }\href {\doibase 10.1038/nature25504}
  {\bibfield  {journal} {\bibinfo  {journal} {Nature}\ }\textbf {\bibinfo
  {volume} {554}},\ \bibinfo {pages} {360} (\bibinfo {year}
  {2018})}\BibitemShut {NoStop}%
\bibitem [{\citenamefont {Bick}\ \emph {et~al.}(2011)\citenamefont {Bick},
  \citenamefont {Timme}, \citenamefont {Paulikat}, \citenamefont {Rathlev},\
  and\ \citenamefont {Ashwin}}]{bick2011}%
  \BibitemOpen
  \bibfield  {author} {\bibinfo {author} {\bibfnamefont {C.}~\bibnamefont
  {Bick}}, \bibinfo {author} {\bibfnamefont {M.}~\bibnamefont {Timme}},
  \bibinfo {author} {\bibfnamefont {D.}~\bibnamefont {Paulikat}}, \bibinfo
  {author} {\bibfnamefont {D.}~\bibnamefont {Rathlev}}, \ and\ \bibinfo
  {author} {\bibfnamefont {P.}~\bibnamefont {Ashwin}},\ }\href {\doibase
  10.1103/PhysRevLett.107.244101} {\bibfield  {journal} {\bibinfo  {journal}
  {Phys. Rev. Lett.}\ }\textbf {\bibinfo {volume} {107}},\ \bibinfo {pages}
  {244101} (\bibinfo {year} {2011})}\BibitemShut {NoStop}%
\bibitem [{\citenamefont {Atsumi}\ and\ \citenamefont
  {Nakao}(2012)}]{atsumi2012}%
  \BibitemOpen
  \bibfield  {author} {\bibinfo {author} {\bibfnamefont {Y.}~\bibnamefont
  {Atsumi}}\ and\ \bibinfo {author} {\bibfnamefont {H.}~\bibnamefont {Nakao}},\
  }\href {\doibase 10.1103/PhysRevE.85.056207} {\bibfield  {journal} {\bibinfo
  {journal} {Phys. Rev. E}\ }\textbf {\bibinfo {volume} {85}},\ \bibinfo
  {pages} {056207} (\bibinfo {year} {2012})}\BibitemShut {NoStop}%
\bibitem [{\citenamefont {Newman}\ \emph {et~al.}(2018)\citenamefont {Newman},
  \citenamefont {Lucas},\ and\ \citenamefont {Stefanovska}}]{newman2018}%
  \BibitemOpen
  \bibfield  {author} {\bibinfo {author} {\bibfnamefont {J.}~\bibnamefont
  {Newman}}, \bibinfo {author} {\bibfnamefont {M.}~\bibnamefont {Lucas}}, \
  and\ \bibinfo {author} {\bibfnamefont {A.}~\bibnamefont {Stefanovska}},\
  }\href@noop {} {\enquote {\bibinfo {title} {Limitations of the asymptotic
  approach to dynamics},}\ } (\bibinfo {year} {2018}),\ \bibinfo {note} {with
  referees}\BibitemShut {NoStop}%
\bibitem [{\citenamefont {Petkoski}\ and\ \citenamefont
  {Stefanovska}(2012)}]{petkoski2012}%
  \BibitemOpen
  \bibfield  {author} {\bibinfo {author} {\bibfnamefont {S.}~\bibnamefont
  {Petkoski}}\ and\ \bibinfo {author} {\bibfnamefont {A.}~\bibnamefont
  {Stefanovska}},\ }\href {\doibase 10.1103/PhysRevE.86.046212} {\bibfield
  {journal} {\bibinfo  {journal} {Phys. Rev. E}\ }\textbf {\bibinfo {volume}
  {86}},\ \bibinfo {pages} {046212} (\bibinfo {year} {2012})}\BibitemShut
  {NoStop}%
\bibitem [{\citenamefont {Pietras}\ and\ \citenamefont
  {Daffertshofer}(2016)}]{pietras2016}%
  \BibitemOpen
  \bibfield  {author} {\bibinfo {author} {\bibfnamefont {B.}~\bibnamefont
  {Pietras}}\ and\ \bibinfo {author} {\bibfnamefont {A.}~\bibnamefont
  {Daffertshofer}},\ }\href {\doibase 10.1063/1.4963371} {\bibfield  {journal}
  {\bibinfo  {journal} {Chaos}\ }\textbf {\bibinfo {volume} {26}},\ \bibinfo
  {pages} {103101} (\bibinfo {year} {2016})}\BibitemShut {NoStop}%
\bibitem [{\citenamefont {Petit}\ \emph {et~al.}(2017)\citenamefont {Petit},
  \citenamefont {Lauwens}, \citenamefont {Fanelli},\ and\ \citenamefont
  {Carletti}}]{petit2017}%
  \BibitemOpen
  \bibfield  {author} {\bibinfo {author} {\bibfnamefont {J.}~\bibnamefont
  {Petit}}, \bibinfo {author} {\bibfnamefont {B.}~\bibnamefont {Lauwens}},
  \bibinfo {author} {\bibfnamefont {D.}~\bibnamefont {Fanelli}}, \ and\
  \bibinfo {author} {\bibfnamefont {T.}~\bibnamefont {Carletti}},\ }\href
  {\doibase 10.1103/PhysRevLett.119.148301} {\bibfield  {journal} {\bibinfo
  {journal} {Phys. Rev. Lett.}\ }\textbf {\bibinfo {volume} {119}},\ \bibinfo
  {pages} {148301} (\bibinfo {year} {2017})}\BibitemShut {NoStop}%
\bibitem [{\citenamefont {Lu}\ and\ \citenamefont {Atay}(2018)}]{lu2018}%
  \BibitemOpen
  \bibfield  {author} {\bibinfo {author} {\bibfnamefont {W.}~\bibnamefont
  {Lu}}\ and\ \bibinfo {author} {\bibfnamefont {F.~M.}\ \bibnamefont {Atay}},\
  }\href {\doibase 10.1137/16M1084390} {\bibfield  {journal} {\bibinfo
  {journal} {SIAM J. Appl. Dyn. Syst.}\ }\textbf {\bibinfo {volume} {17}},\
  \bibinfo {pages} {457} (\bibinfo {year} {2018})}\BibitemShut {NoStop}%
\bibitem [{\citenamefont {Lucas}\ \emph
  {et~al.}(2018{\natexlab{a}})\citenamefont {Lucas}, \citenamefont {Fanelli},
  \citenamefont {Carletti},\ and\ \citenamefont {Petit}}]{lucas2018b}%
  \BibitemOpen
  \bibfield  {author} {\bibinfo {author} {\bibfnamefont {M.}~\bibnamefont
  {Lucas}}, \bibinfo {author} {\bibfnamefont {D.}~\bibnamefont {Fanelli}},
  \bibinfo {author} {\bibfnamefont {T.}~\bibnamefont {Carletti}}, \ and\
  \bibinfo {author} {\bibfnamefont {J.}~\bibnamefont {Petit}},\ }\href
  {\doibase 10.1209/0295-5075/121/50008} {\bibfield  {journal} {\bibinfo
  {journal} {Europhys. Lett.}\ }\textbf {\bibinfo {volume} {121}},\ \bibinfo
  {pages} {50008} (\bibinfo {year} {2018}{\natexlab{a}})}\BibitemShut {NoStop}%
\bibitem [{\citenamefont {Radicchi}\ and\ \citenamefont
  {Meyer-Ortmanns}(2006)}]{radicchi2006}%
  \BibitemOpen
  \bibfield  {author} {\bibinfo {author} {\bibfnamefont {F.}~\bibnamefont
  {Radicchi}}\ and\ \bibinfo {author} {\bibfnamefont {H.}~\bibnamefont
  {Meyer-Ortmanns}},\ }\href {\doibase 10.1103/PhysRevE.73.036218} {\bibfield
  {journal} {\bibinfo  {journal} {Phys. Rev. E}\ }\textbf {\bibinfo {volume}
  {73}},\ \bibinfo {pages} {036218} (\bibinfo {year} {2006})}\BibitemShut
  {NoStop}%
\bibitem [{\citenamefont {Kori}\ and\ \citenamefont
  {Mikhailov}(2004)}]{kori2004}%
  \BibitemOpen
  \bibfield  {author} {\bibinfo {author} {\bibfnamefont {H.}~\bibnamefont
  {Kori}}\ and\ \bibinfo {author} {\bibfnamefont {A.~S.}\ \bibnamefont
  {Mikhailov}},\ }\href {\doibase 10.1103/PhysRevLett.93.254101} {\bibfield
  {journal} {\bibinfo  {journal} {Phys. Rev. Lett.}\ }\textbf {\bibinfo
  {volume} {93}},\ \bibinfo {pages} {254101} (\bibinfo {year}
  {2004})}\BibitemShut {NoStop}%
\bibitem [{\citenamefont {Kori}\ and\ \citenamefont
  {Mikhailov}(2006)}]{kori2006}%
  \BibitemOpen
  \bibfield  {author} {\bibinfo {author} {\bibfnamefont {H.}~\bibnamefont
  {Kori}}\ and\ \bibinfo {author} {\bibfnamefont {A.~S.}\ \bibnamefont
  {Mikhailov}},\ }\href {\doibase 10.1103/PhysRevE.74.066115} {\bibfield
  {journal} {\bibinfo  {journal} {Phys. Rev. E}\ }\textbf {\bibinfo {volume}
  {74}},\ \bibinfo {pages} {66115} (\bibinfo {year} {2006})}\BibitemShut
  {NoStop}%
\bibitem [{\citenamefont {Lucas}\ \emph
  {et~al.}(2018{\natexlab{b}})\citenamefont {Lucas}, \citenamefont {Newman},\
  and\ \citenamefont {Stefanovska}}]{lucas2018}%
  \BibitemOpen
  \bibfield  {author} {\bibinfo {author} {\bibfnamefont {M.}~\bibnamefont
  {Lucas}}, \bibinfo {author} {\bibfnamefont {J.}~\bibnamefont {Newman}}, \
  and\ \bibinfo {author} {\bibfnamefont {A.}~\bibnamefont {Stefanovska}},\
  }\href {\doibase 10.1103/PhysRevE.97.042209} {\bibfield  {journal} {\bibinfo
  {journal} {Phys. Rev. E}\ }\textbf {\bibinfo {volume} {97}},\ \bibinfo
  {pages} {042209} (\bibinfo {year} {2018}{\natexlab{b}})}\BibitemShut
  {NoStop}%
\bibitem [{\citenamefont {Kurebayashi}\ \emph {et~al.}(2013)\citenamefont
  {Kurebayashi}, \citenamefont {Shirasaka},\ and\ \citenamefont
  {Nakao}}]{kurebayashi2013}%
  \BibitemOpen
  \bibfield  {author} {\bibinfo {author} {\bibfnamefont {W.}~\bibnamefont
  {Kurebayashi}}, \bibinfo {author} {\bibfnamefont {S.}~\bibnamefont
  {Shirasaka}}, \ and\ \bibinfo {author} {\bibfnamefont {H.}~\bibnamefont
  {Nakao}},\ }\href {\doibase 10.1103/PhysRevLett.111.214101} {\bibfield
  {journal} {\bibinfo  {journal} {Phys. Rev. Lett.}\ }\textbf {\bibinfo
  {volume} {111}},\ \bibinfo {pages} {214101} (\bibinfo {year}
  {2013})}\BibitemShut {NoStop}%
\bibitem [{\citenamefont {Kurebayashi}\ \emph {et~al.}(2015)\citenamefont
  {Kurebayashi}, \citenamefont {Shirasaka},\ and\ \citenamefont
  {Nakao}}]{kurebayashi2015}%
  \BibitemOpen
  \bibfield  {author} {\bibinfo {author} {\bibfnamefont {W.}~\bibnamefont
  {Kurebayashi}}, \bibinfo {author} {\bibfnamefont {S.}~\bibnamefont
  {Shirasaka}}, \ and\ \bibinfo {author} {\bibfnamefont {H.}~\bibnamefont
  {Nakao}},\ }\href {\doibase 10.1587/nolta.6.171} {\bibfield  {journal}
  {\bibinfo  {journal} {NOLTA}\ }\textbf {\bibinfo {volume} {6}},\ \bibinfo
  {pages} {171} (\bibinfo {year} {2015})}\BibitemShut {NoStop}%
\bibitem [{\citenamefont {Park}\ and\ \citenamefont
  {Ermentrout}(2016)}]{park2016}%
  \BibitemOpen
  \bibfield  {author} {\bibinfo {author} {\bibfnamefont {Y.}~\bibnamefont
  {Park}}\ and\ \bibinfo {author} {\bibfnamefont {G.~B.}\ \bibnamefont
  {Ermentrout}},\ }\href {\doibase 10.1007/s10827-016-0596-6} {\bibfield
  {journal} {\bibinfo  {journal} {J. Comput. Neurosci.}\ }\textbf {\bibinfo
  {volume} {40}},\ \bibinfo {pages} {269} (\bibinfo {year} {2016})}\BibitemShut
  {NoStop}%
\bibitem [{\citenamefont {Toenjes}\ and\ \citenamefont
  {Blasius}(2009)}]{toenjes2009}%
  \BibitemOpen
  \bibfield  {author} {\bibinfo {author} {\bibfnamefont {R.}~\bibnamefont
  {Toenjes}}\ and\ \bibinfo {author} {\bibfnamefont {B.}~\bibnamefont
  {Blasius}},\ }\href {\doibase 10.1103/PhysRevE.80.026202} {\bibfield
  {journal} {\bibinfo  {journal} {Phys. Rev. E}\ }\textbf {\bibinfo {volume}
  {80}},\ \bibinfo {pages} {026202} (\bibinfo {year} {2009})}\BibitemShut
  {NoStop}%
\bibitem [{\citenamefont {Jensen}(2002)}]{jensen2002}%
  \BibitemOpen
  \bibfield  {author} {\bibinfo {author} {\bibfnamefont {R.}~\bibnamefont
  {Jensen}},\ }\href {\doibase 10.1119/1.1467909} {\bibfield  {journal}
  {\bibinfo  {journal} {Am. J. Phys.}\ }\textbf {\bibinfo {volume} {70}},\
  \bibinfo {pages} {607} (\bibinfo {year} {2002})}\BibitemShut {NoStop}%
\bibitem [{\citenamefont {Almendral}\ and\ \citenamefont
  {D{\'\i}az-Guilera}(2007)}]{almendral2007}%
  \BibitemOpen
  \bibfield  {author} {\bibinfo {author} {\bibfnamefont {J.~A.}\ \bibnamefont
  {Almendral}}\ and\ \bibinfo {author} {\bibfnamefont {A.}~\bibnamefont
  {D{\'\i}az-Guilera}},\ }\href {\doibase 10.1088/1367-2630/9/6/187} {\bibfield
   {journal} {\bibinfo  {journal} {New J. Phys.}\ }\textbf {\bibinfo {volume}
  {9}},\ \bibinfo {pages} {187} (\bibinfo {year} {2007})}\BibitemShut {NoStop}%
\bibitem [{Note1()}]{Note1}%
  \BibitemOpen
  \bibinfo {note} {The network topology was created with the Python function
  \protect \texttt {barabasi\protect \_albert\protect \_graph(N, m)} from the
  \protect \texttt {NetworkX} package~\cite {scipy_proceedings11}, with $N=30$
  the total number of nodes, and $m=3$ the number of nodes added.}\BibitemShut
  {Stop}%
\bibitem [{\citenamefont {Ko}\ and\ \citenamefont {Ermentrout}(2008)}]{ko2008}%
  \BibitemOpen
  \bibfield  {author} {\bibinfo {author} {\bibfnamefont {T.-W.}\ \bibnamefont
  {Ko}}\ and\ \bibinfo {author} {\bibfnamefont {G.~B.}\ \bibnamefont
  {Ermentrout}},\ }\href {\doibase 10.1103/PhysRevE.78.016203} {\bibfield
  {journal} {\bibinfo  {journal} {Phys. Rev. E}\ }\textbf {\bibinfo {volume}
  {78}},\ \bibinfo {pages} {16203} (\bibinfo {year} {2008})}\BibitemShut
  {NoStop}%
\bibitem [{\citenamefont {Kuramoto}\ and\ \citenamefont
  {Battogtokh}(2002)}]{kuramoto2002}%
  \BibitemOpen
  \bibfield  {author} {\bibinfo {author} {\bibfnamefont {Y.}~\bibnamefont
  {Kuramoto}}\ and\ \bibinfo {author} {\bibfnamefont {D.}~\bibnamefont
  {Battogtokh}},\ }\href@noop {} {\bibfield  {journal} {\bibinfo  {journal}
  {Nonlinear Phenom. Complex Systems}\ }\textbf {\bibinfo {volume} {5}},\
  \bibinfo {pages} {380} (\bibinfo {year} {2002})}\BibitemShut {NoStop}%
\bibitem [{\citenamefont {Panaggio}\ and\ \citenamefont
  {Abrams}(2015)}]{panaggio2015}%
  \BibitemOpen
  \bibfield  {author} {\bibinfo {author} {\bibfnamefont {M.~J.}\ \bibnamefont
  {Panaggio}}\ and\ \bibinfo {author} {\bibfnamefont {D.~M.}\ \bibnamefont
  {Abrams}},\ }\href {\doibase 10.1088/0951-7715/28/3/R67} {\bibfield
  {journal} {\bibinfo  {journal} {Nonlinearity}\ }\textbf {\bibinfo {volume}
  {28}},\ \bibinfo {pages} {R67} (\bibinfo {year} {2015})}\BibitemShut
  {NoStop}%
\bibitem [{\citenamefont {Hagberg}\ \emph {et~al.}(2008)\citenamefont
  {Hagberg}, \citenamefont {Schult},\ and\ \citenamefont
  {Swart}}]{scipy_proceedings11}%
  \BibitemOpen
  \bibfield  {author} {\bibinfo {author} {\bibfnamefont {A.~A.}\ \bibnamefont
  {Hagberg}}, \bibinfo {author} {\bibfnamefont {D.~A.}\ \bibnamefont {Schult}},
  \ and\ \bibinfo {author} {\bibfnamefont {P.~J.}\ \bibnamefont {Swart}},\ }in\
  \href@noop {} {\emph {\bibinfo {booktitle} {Proceedings of the 7th Python in
  Science Conference}}}\ (\bibinfo {address} {Pasadena, CA USA},\ \bibinfo
  {year} {2008})\ pp.\ \bibinfo {pages} {11--15}\BibitemShut {NoStop}%
\end{thebibliography}%

\end{document}